\documentstyle[12pt]{article}

\begin{document}

\vspace*{1cm}
\begin{center}
{\large Study of $\Lambda$ hypernuclei in the quark mean field model\\}

\vspace*{1.5cm}
{H. Shen \\
{\it CCAST(World Lab.), P.O.Box 8730, Beijing 100080, China\\
Department of Physics, Nankai University, Tianjin 300071, China\\}}

\vspace*{0.5cm}
{H. Toki \\
{\it Research Center for Nuclear Physics (RCNP), Osaka
University,\\ Ibaraki, Osaka 567-0047, Japan\\}}
\end{center}

\vspace*{1cm}

\begin{abstract}
We extend the quark mean field model to the study of $\Lambda$
hypernuclei. Without adjusting parameters, the properties of
$\Lambda$ hypernuclei can be described reasonably well. The small
spin-orbit splittings for $\Lambda$ in hypernuclei are achieved,
while the $\Lambda$ single particle energies in the present model
are slightly underestimated as compared with the experimental
values. About $3\%$ deviation from the quark model prediction for
the $\omega-\Lambda$ couplings is required in order to reproduce
the experimental single particle energies.

\vspace*{1cm}


\end{abstract}
\newpage

Over the past years hypernuclear physics has been attracting great
interest. The most extensively studied hypernuclear system
consists of a single $\Lambda$ particle coupled to the nuclear
core, which could be produced in ($\pi^+$,$K^+$) or
($K^-$,$\pi^-$) reactions. $\Lambda$ hypernuclear spectroscopy
indicates that $\Lambda$ is weakly bound in nuclear medium and its
spin-orbit splitting is quite small compared with the
nucleon~\cite{exp1,exp2,exp3,exp4}. Theoretical efforts have been
devoted to understanding the properties of
hypernuclei~\cite{rmf1}. The relativistic mean field (RMF) models
have been successfully applied to describe hypernuclei with
adjustable meson-hyperon couplings and tensor
couplings~\cite{rmf2,rmf3,rmf4,rmf5}. Microscopic hypernuclear
structure calculations have also been performed in the quark-meson
coupling (QMC) model~\cite{qmc1,qmc2}, where both the hyperon and
the nucleon are the composites of quarks.

In this paper, $\Lambda$ hypernuclei are investigated in the
quark mean field (QMF) model, which has been successfully applied
to study the properties of both nuclear matter and finite
nuclei~\cite{qmf1,qmf2}.
The QMF model describes the nucleon in terms of the constituent
quark model, while the MIT bag model was used in the QMC
model~\cite{qmc3,qmc4,qmc5}.
The main purpose of the present work is to extend the QMF model
to flavor SU(3) and further to the study of $\Lambda$ hypernuclei.
With the assumption that the $\sigma$, $\omega$, and $\rho$
mesons couple only to the $u$ and $d$ quarks, the meson-hyperon
couplings can be obtained from the quark model.
It is very interesting to perform the self-consistent calculations
for $\Lambda$ hypernuclei in the QMF model without any freedom
of adjusting parameters.

We start with the description of the $\Lambda$ hyperon in nuclear
medium. The nucleon and the $\Lambda$ hyperon as composites of
three quarks are described in terms of the constituent quark model,
in which the constituent quarks satisfy the Dirac equations
with confinement potentials~\cite{qmf2}.
The $\sigma$ meson, which couples directly to the $u$ and $d$ quarks,
provides a scalar potential to the quark and as a
consequence reduces the constituent quark mass.
The change of the nucleon properties under the influence
of the $\sigma$ mean field has been studied in Ref.~\cite{qmf2}.
We now treat the $\Lambda$ hyperon in nuclear medium.
According to the OZI rule, the non-strange mesons
couple exclusively to the $u$ and $d$ quarks and not to the $s$ quark.
Therefore the state of the $s$ quark inside the $\Lambda$ hyperon
will not be influenced due to the presence of the non-strange meson
mean fields. We follow Ref.~\cite{qmf2} to take into account the spin correlations and remove the spurious center of mass motion,
and then obtain the effective mass for $\Lambda$ hyperon as
$M_{\Lambda}^{*}=\sqrt{(2e_q+e_s+E^{\Lambda}_{\rm{spin}})^2-
                 (2\langle p_q^2 \rangle+\langle p_s^2 \rangle)}$,
while the effective mass for nucleon is expressed as
$M_N^{*}=\sqrt{(3e_q+E^N_{\rm{spin}})^2-
                 3\langle p_q^2 \rangle }$.
Here the subscript $q$ denotes the $u$ or $d$ quark. The energies
($e_q$ and $e_s$) and momenta ($\langle p_q^2 \rangle$ and
$\langle p_s^2 \rangle$) can be obtained by solving the Dirac
equations. We take the same two types of confinement as used in
Ref.~\cite{qmf2}: (1) scalar potential $\chi_c=\frac{1}{2}kr^2$
and (2) scalar-vector potential $\chi_c=\frac{1}{2}kr^2
(1+\gamma^0)/2$ with $k=700 \;\rm{MeV/fm^2}$. The quark masses are
taken as $m_q=313\;\rm{MeV}$ and $m_s=490\;\rm{MeV}$. The spin
correlations ($E^N_{\rm{spin}}$ and $E^{\Lambda}_{\rm{spin}}$) are
fixed by fitting the nucleon and $\Lambda$ masses in free space
($M_N=939\;\rm{MeV}$, $M_{\Lambda}=1116\;\rm{MeV}$).

We present in Fig. 1 the variations of the effective masses for
the nucleon and the $\Lambda$ hyperon, $\delta M_i^*=M^*_i-M_i$
($i=N,\Lambda$), as functions of the quark mass correction due to
the presence of the $\sigma$ mean field, $\delta
m_q=m_q-m_q^{*}=-g_{\sigma}^q\sigma$ ($q=u,d$). The results with
the scalar potential are shown by solid curves, while those with
the scalar-vector potential by dashed ones. The behavior of the
effective nucleon mass has been discussed extensively in our
previous work~\cite{qmf2}. We now focus on the effective mass for
the $\Lambda$ hyperon. It is obvious that the reduction of
$M_{\Lambda}^{*}$ is smaller than that of $M_N^*$, since only two
of the three quarks in the $\Lambda$ hyperon are influenced by the
$\sigma$ mean field. We note that the dependence of the effective
masses on the $\sigma$ mean field must be calculated
self-consistently within the quark model, therefore the ratio of
the variation of the effective mass for the $\Lambda$ hyperon to
that for the nucleon, $\delta M_{\Lambda}^{*} / \delta M_N^{*}$,
is not so simple as a constant as in the RMF
models~\cite{rmf2,rmf3}.

We treat a single $\Lambda$ hypernucleus as a system of many nucleons and
a $\Lambda$ hyperon which interact through exchange of
$\sigma$, $\omega$, and $\rho$ mesons.
The effective Lagrangian can be written as
\begin{eqnarray}
{\cal L} &=&
\bar\psi\left[ i\gamma_\mu\partial^\mu-M_N^*
-g_\omega \omega \gamma^0
-g_\rho \rho \tau_3\gamma^0
-e\frac{(1+\tau_3)}{2} A \gamma^0
\right] \psi  \\ \nonumber
 & &
+\bar\psi_{\Lambda} \left[ i\gamma_\mu\partial^\mu-M_{\Lambda}^*
-g^{\Lambda}_\omega \omega \gamma^0
\right] \psi_{\Lambda} \\ \nonumber
 & &
-\frac{1}{2} (\bigtriangledown\sigma)^2
-\frac{1}{2} m_\sigma^2\sigma^2
-\frac{1}{4} g_3\sigma^4
+\frac{1}{2} (\bigtriangledown\omega)^2
+\frac{1}{2} m_\omega^2\omega^2
+\frac{1}{4} c_3\omega^4   \\ \nonumber
 & &
+\frac{1}{2} (\bigtriangledown\rho)^2
+\frac{1}{2} m_\rho^2\rho^2
+\frac{1}{2}(\bigtriangledown A)^2,
\end{eqnarray}
where $\psi$ and $\psi_{\Lambda}$ are the Dirac spinors for the
nucleon and the $\Lambda$ hyperon. The mean field approximation
has been adopted for the exchanged $\sigma$, $\omega$, and $\rho$
mesons, while the mean field values of these mesons are denoted by
$\sigma$, $\omega$, and $\rho$, respectively. $m_{\sigma}$,
$m_{\omega}$, and $m_{\rho}$ are the meson masses. $A$ is the
electromagnetic field which couples to the protons. Since the
$\Lambda$ hyperon is neutral and isoscalar, it only couples to the
$\sigma$ and $\omega$ mesons. The influence of the $\sigma$ meson
on the $\Lambda$ hyperon is contained in $M_{\Lambda}^{*}$, while
the $\omega$ meson couples to the $\Lambda$ hyperon with the
coupling constant $g^{\Lambda}_\omega=2g_\omega^q$ (for the
nucleon, $g_{\omega}=3g_\omega^q$, $g_\rho=g_\rho^q$). In the QMF
model, the basic parameters are the quark-meson couplings
($g^q_\sigma$, $g_\omega^q$, and $g_\rho^q$), the nonlinear
self-coupling constants ($g_3$ and $c_3$), and the mass of the
$\sigma$ meson ($m_\sigma$), which has been determined by fitting
the properties of nuclear matter and finite nuclei in
Ref.~\cite{qmf2}. Therefore, no more adjustable parameters exist
when it is extended to the calculation of $\Lambda$ hypernuclei.
From the Lagrangian given in (1), we obtain the following
Euler-Lagrange equations
\begin{eqnarray}
 & & \left[
i\gamma_{\mu}\partial^{\mu}-M_N^*
-g_\omega \omega \gamma^0
-g_\rho \rho \tau_3\gamma^0\
-e\frac{(1+\tau_3)}{2} A \gamma^0
 \right]\psi
= 0, \\
 & & \left[
i\gamma_{\mu}\partial^{\mu}-M_\Lambda^*
-g^\Lambda_\omega \omega \gamma^0
 \right]\psi_\Lambda
= 0,
\\
 & & \left(
 -\Delta+m_\sigma^2\right) \sigma=
-\frac {\partial M_N^*}{\partial \sigma} \rho_s
-\frac {\partial M_\Lambda^*}{\partial \sigma} \rho^\Lambda_s
-g_3 \sigma^3,
\\
 & & \left(
 -\Delta+m_\omega^2\right) \omega=
g_\omega \rho_v +g^\Lambda_\omega \rho^\Lambda_v
-c_3 \omega^3,
\\
 & & \left(
 -\Delta+m_\rho^2\right) \rho =
g_\rho \rho_3,
\\
 & &
 -\Delta A =
e \rho_p,
\end{eqnarray}
where $\rho_s$ ($\rho_s^\Lambda$), $\rho_v$ ($\rho_v^\Lambda$),
      $\rho_3$, and $\rho_p$
are the scalar, vector, third component of isovector, and proton
densities, respectively. The above coupled equations are solved
self-consistently with the effective masses ($M_N^{*}$ and
$M_{\Lambda}^{*}$) obtained at the quark level.

If we consider the spin of $\Lambda$ is carried exclusively by the
$s$ quark, the spin-orbit interaction should come entirely from
the Thomas precession~\cite{qmc1,qmc3}. The correction,
$\frac{-1}{{M_\Lambda^*}^2\;r} \left( \frac{d}{dr}
g_\omega^\Lambda\omega \right)\; \vec l \cdot \vec s $, could be
added perturbatively to the $\Lambda$ single particle energies, so
that the spin-orbit interaction corresponds to the formula from
Thomas precession~\cite{qmc1}. With this correction, small
spin-orbit splittings are obtained while the single particle
energies are not much altered. For instance, the spin-orbit
splitting for the $1f$ states ($1f_{5/2}-1f_{7/2}$) in
$^{91}_\Lambda\rm{Zr}$ decreases to $0.03\;\rm{MeV}$, while the
value is $1.5\;\rm{MeV}$ without the perturbative correction. The
recent experimental result~\cite{exp4} seems to reveal the
splitting of $1.6\pm 0.15\;\rm{MeV}$ for the $1f$ states in
$^{89}_\Lambda\rm{Y}$.

We present in Fig. 2 the calculated $\Lambda$ single particle
energies in several hypernuclei consisting of a closed-shell
nuclear core and a single $\Lambda$ hyperon,
while the results in the QMC model~\cite{qmc1}
and the experimental values~\cite{exp1,exp2,exp3}
are also shown for comparison.
Here the QMC results do not contain the effect of the Pauli blocking,
which has been included phenomenologically in Ref.~\cite{qmc2}
in order to reproduce the experimental single particle energies.
The $\rm{QMF^I}$ and $\rm{QMF^{II}}$ denote the models with
confinements
    $\chi_c=\frac{1}{2}kr^2 (1+\gamma^0)/2$ and
    $\chi_c=\frac{1}{2}kr^2$, respectively.
The parameters in the QMF models have been determined
in our previous work~\cite{qmf2}, therefore no free
parameter in the present calculation.
We notice that the final results are insensitive to the choice of
the $s$ quark mass.
It is found that small spin-orbit splittings for the $\Lambda$
in those hypernuclei are obtained in the present model.
For instance, the spin-orbit splittings for the $1d$ states
($1d_{3/2}-1d_{5/2}$) in $^{41}_\Lambda\rm{Ca}$
and $^{209}_\Lambda\rm{Pb}$ decrease to
$0.02$ and $0.01\;\rm{MeV}$ with the perturbative correction,
while those values are about $1.4$ and $0.5\;\rm{MeV}$
without the perturbative correction.
The small spin-orbit splittings were mostly worked out
by adding tensor interactions in the RMF models~\cite{rmf2,rmf3,rmf4}.
In Fig. 3, we plot the scalar and vector potentials
($U_S$ and $U_V$) for the $1s_{1/2}$ $\Lambda$
state in $^{41}_\Lambda\rm{Ca}$ and $^{209}_\Lambda\rm{Pb}$.
The results with two types of confinement
shown by solid and dashed curves are almost identical.
The attractive scalar potential is mostly canceled by the
repulsive vector potential, and their difference at the center
of the hypernuclei is about $20-25\; \rm{MeV}$.

The single particle energies in the present model seem to be slightly
underestimated, which is opposite to the tendency in the QMC model.
It is well known that the properties of $\Lambda$ hypernuclei are very
sensitive to the effective coupling constants on the hadronic
level, especially the two relative couplings
    $R_\sigma=g^\Lambda_\sigma/g_\sigma$
and $R_\omega=g^\Lambda_\omega/g_\omega$~\cite{rmf6}. The quark
model value, $R_\sigma=R_\omega=2/3$, usually gives large
overbinding of $\Lambda$ single particle energies. Some effects,
like correlated $\pi\pi$ exchange, may cause the deviations of
$R_\sigma$ and $R_\omega$ from the quark model value of $2/3$.
Most studies of the hypernuclei in the RMF models are performed by
treating both $R_\sigma$ and $R_\omega$ (or only one of them) as
phenomenological parameters, which are fitted by using
experimental data~\cite{rmf2,rmf3,rmf4,rmf6}. In the present
model, $R_\sigma=g^\Lambda_\sigma/g_\sigma
  =\left[\frac{\partial M_\Lambda^*}{\partial \sigma}\right]
  /\left[\frac{\partial       M_N^*}{\partial \sigma}\right]$
must be calculated self-consistently on the quark level, while
$R_\omega=2/3$ is based on the quark model. Comparing with the RMF
models, $R_\sigma$ in the QMF model depends on the $\sigma$ mean
field, and could not be a constant again. The resulting $\Lambda$
single particle energies are slightly underestimated in comparison
with the experimental values as shown in Fig. 2. The results can
be largely improved if we use the scaled coupling constant, $0.97
\times g_\omega^{\Lambda}$, which gives the $1s_{1/2}$ single
particle energy in $^{209}_\Lambda\rm{Pb}$ to be $-27.2\;\rm{MeV}$
in $\rm{QMF^I}$ and $-26.0\;\rm{MeV}$ in $\rm{QMF^{II}}$. On the
contrary, the results in the QMC model ( without Pauli blocking
effect ) are overestimated in comparison with the experimental
data~\cite{qmc1,qmc2}. The scaled coupling constant, $1.10 \times
g_\omega^{\Lambda}$ ( or $0.93 \times g_\sigma^{\Lambda}$ ), is
required in order to reproduce the experimental single particle
energies in the QMC model.

It is very interesting to discuss the origin of this difference,
because of the similarity of the QMC and QMF models.
$R_\omega=2/3$ has been used in both models, while $R_\sigma$ has
to be calculated with different approaches. $R_\sigma$ obtained in
the QMC model was very close to $2/3$. This is related to its
expressions for the effective masses, where the center of mass
correction was assumed to be independent of the $\sigma$ mean
field, and parametrized into the term of $Z_0$. In our
calculation, we keep the center of mass correction, which is found
to decrease with increasing $\sigma$ mean field, and then we get
smaller values than $2/3$ for $R_\sigma$. This seems to be the
dominant origin of the underbinding in the QMF model, to be
contrasted with the overbinding in the QMC model.

In summary, we have reported the results of the first application
of the QMF model to the description of $\Lambda$ hypernuclei.
With the parameters determined by the properties
of nuclear matter and finite nuclei~\cite{qmf2},
the calculated results for $\Lambda$ hypernuclei are acceptable.
The small spin-orbit splittings for the $\Lambda$ in hypernuclei
are obtained in the present model, while the single particle energies
are slightly underestimated in comparison with the experimental values.
We notice that about $3\%$ deviation from $R_\omega=2/3$ is required in
order to reproduce the experimental single particle energies.
The $3\%$ reduction in $R_\omega$ provides $\sim 6\; \rm{MeV}$
less repulsion, so that the $\Lambda$ potential decreases to
the empirical value ( $U_\Lambda\approx 25-30\; \rm{MeV}$ ).
In the present calculation,
the effective masses for the nucleon and the $\Lambda$ hyperon,
which play important roles in getting the final results,
are obtained with the assumptions that
the spin correlations and the confining potentials are not modified
in nuclear matter. It is a challenging work to study the variations
of those values in nuclear medium.

\section*{Acknowledgments}
This work was supported in part by the National Natural Science
Foundation of China.


\section*{Figure captions}

\begin{description}

 \item[Figure 1:]
    The variations of the effective masses for the nucleon and
    the $\Lambda$ hyperon, $\delta M_i^*=M^*_i-M_i$ ($i=N,\Lambda$),
    as functions of the quark mass correction, $\delta m_q$.
    The results in the QMF model with
    $\chi_c=\frac{1}{2}kr^2$ are shown by solid curves,
    while those with $\chi_c=\frac{1}{2}kr^2 (1+\gamma^0)/2$
    are shown by dashed curves.

 \item[Figure 2:]
    $\Lambda$ Single particle energies in $^{41}_\Lambda\rm{Ca}$,         $^{91}_\Lambda\rm{Zr}$, and $^{209}_\Lambda\rm{Pb}$.
    $\rm{QMF^I}$ and $\rm{QMF^{II}}$ denote the models with
    $\chi_c=\frac{1}{2}kr^2 (1+\gamma^0)/2$ and
    $\chi_c=\frac{1}{2}kr^2$, respectively.
    The results in the QMC model~\protect\cite{qmc1}
    are also shown for comparison.
    The experimental data are taken from
    Refs.~\protect\cite{exp1,exp2,exp3}.

\item[Figure 3:]
    The scalar and vector potentials,
    $U_S$ and $U_V$, for the $1s_{1/2}$ $\Lambda$
    state in $^{41}_\Lambda\rm{Ca}$ and $^{209}_\Lambda\rm{Pb}$.

\end{description}

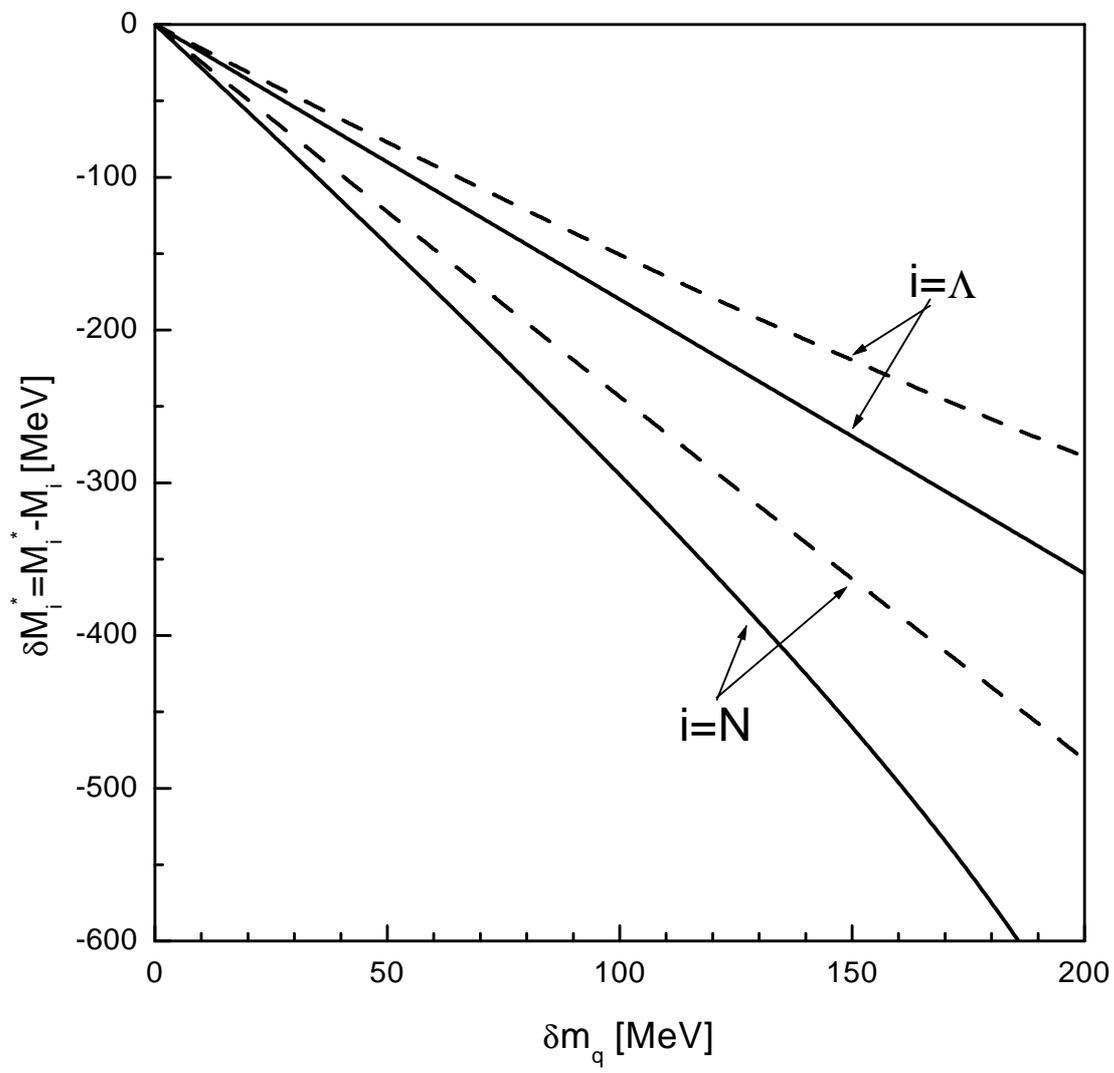
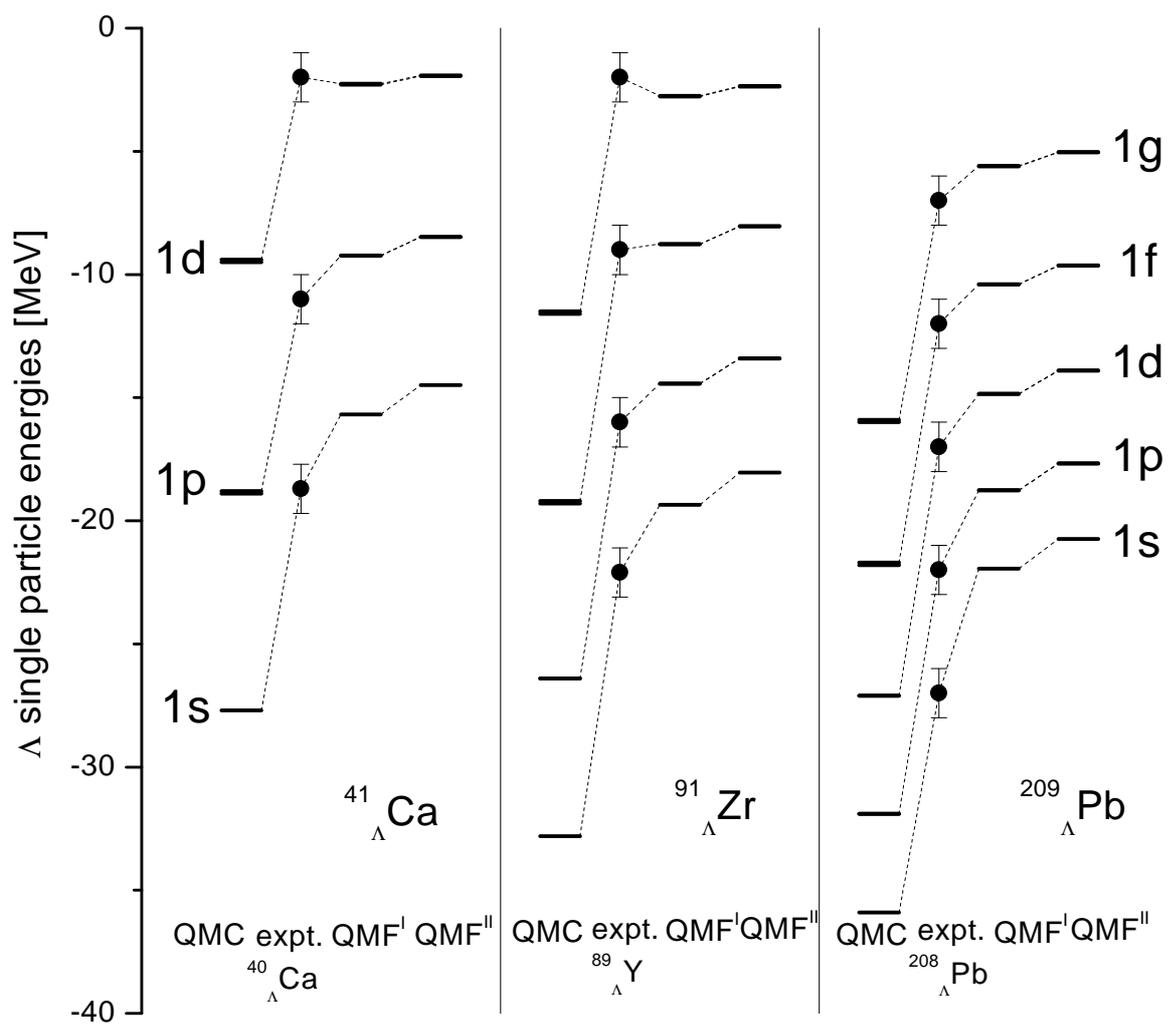
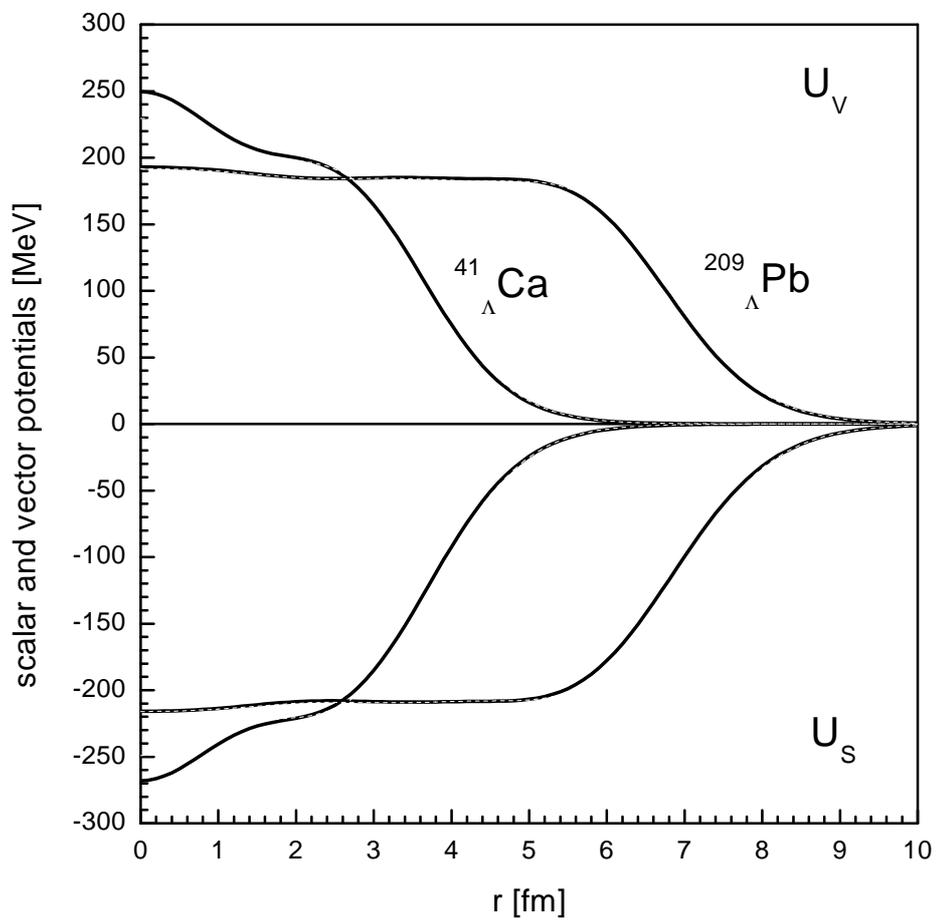
\end{document}